\documentclass[review]{elsarticle}
	\usepackage{lineno,hyperref}
	\journal{Journal of \LaTeX\ Templates}
	\usepackage[russian,english]{babel}
	\usepackage{amsthm}
	\theoremstyle{plain}

	\newtheorem{theorem}{Theorem}
	\newtheorem{definition}{Definition}
	\newtheorem{lemma}{Lemma}
	\newtheorem{corollary}{Corollary}
	\newtheorem*{proof*}{Proof}
	\usepackage{lineno,hyperref}
	\usepackage{ucs} 
	\usepackage[russian]{babel}  % Включаем пакет для поддержки русского языка 
	\usepackage{tikz} 
	
\begin{document}

		\begin{frontmatter}
			
			\title{Application of Analytic Functions to the Global Solvabilty of the Cauchy Problem for Equations of Viscous Incompressible Liquid}
			%\tnotetext[mytitlenote]{Fully documented templates are available in the elsarticle package on %\href{http://www.ctan.org/tex-archive/macros/latex/contrib/elsarticle}{CTAN}.}
			
			\author{Durmagambetov A.A\fnref{myfootnote}}
			\address{}
			\fntext[myfootnote]{}
			\ead[url]{}
			
			\ead{aset.durmagambet@gmail.com}
			
			\address[mymainaddress]{010000, Kazakhstan }
			
			\begin{abstract}
				Using the example of a complicated problem such as the Cauchy problem for the Navier--Stokes equation, we show how the Poincar\'e--Riemann--Hilbert boundary-value problem enables us to construct effective estimates of solutions for this case. The apparatus of the three-dimensional inverse problem of quantum scattering theory is developed for this. 
			\end{abstract}
			
			
		\end{frontmatter}

%	\maketitle
	
	\section {Introduction}
	Using the example of a complicated problem such as the Cauchy problem for the Navier--Stokes equation, we show how the Poincar\'e--Riemann--Hilbert boundary-value problem enables us to construct effective estimates of solutions for this case. The apparatus of the three-dimensional inverse problem of quantum scattering theory is developed for this. It is shown that the unitary scattering operator can be studied as a solution of the Poincar\'e--Riemann--Hilbert boundary-value problem. This allows us to go on to study the potential in the Schr\"odinger equation, which we consider as a velocity component in the Navier--Stokes equation. The same scheme of reduction of Riemann integral equations for the zeta function to the Poincar\'e--Riemann--Hilbert boundary-value problem allows us to construct effective estimates that describe the behaviour of the zeros of the zeta function very well.

	\section{Results for the one-dimensional case}
	%\subsection{General}
	Let us consider a one-dimensional function $ {f}  $ and its Fourier transformation $  \tilde{f} $. Using the notions of module and phase, we write the Fourier transformation in the following form: $  \tilde{f}=|\tilde{f}|\exp(i\Psi) $ , where $ \Psi $ is the phase.  The Plancherel equality states that $ ||f||_{L_{2}}={\rm const}||\tilde{f} ||_ {L_{2}} $. Here we can see that the phase does not contribute to determination of the $X$ norm. To estimate the maximum we make a simple estimate as $  {\rm max}|f|^2 \le 2||f||_{L_{2}} ||\nabla f||_{L_{2}}   $. Now we have an estimate of the function maximum in which the phase is not involved. Let us consider the behaviour of a progressing wave travelling with a constant velocity of $  v=a$ described by the function $ {F(x,t)=f(x+at)}  $.  Its Fourier transformation with respect to the variable $x$ is $  \tilde{F}=\tilde{f}{\exp}(iatk) $.  Again, in this case, we can see that when we study a module of the Fourier transformation, we will not obtain major physical information about the wave, such as its velocity and location of the wave crest because $ |\tilde{F}|=|\tilde{f}| $ . These two examples show the weaknesses of studying the Fourier transformation. Many researchers focus on the study of functions using the embedding theorem, in which the main object of the study is the module of the function. However, as we have seen in the given examples, the phase is a principal physical characteristic of any process, and as we can see in mathematical studies that use the embedding theorem with energy estimates, the phase disappears. Along with the phase, all reasonable information about the physical process disappears, as demonstrated by Tao [1] and other research studies. In fact, Tao built progressing waves that  are  not followed by energy estimates . Let us proceed with a more essential analysis of the influence of the phase on the behaviour of functions.
	\begin  {theorem}
	There are functions of $ W_{2}^{1}(R)$ with a constant rate of the norm for a gradient catastrophe for which a phase change of its Fourier transformation is sufficient.
	\end  {theorem}
	Proof: To prove this, we consider a sequence of testing functions $  \tilde{f_{n}}=\Delta/(1+k^2),\;\Delta=(i-k)^n/(i+k)^n $. It is obvious that $|\tilde{f_{n}}|=1/(1+k^2)$ and
	$  {\rm max}|f_{n}|^2 \le 2||f_{n}||_{L_{2}} ||\nabla f_{n}||_{L_{2}} \le {\rm const}   $. Calculating the Fourier transformation of these testing functions, we obtain 
	\begin{equation}
	f_{n}(x) =	x(-1)^{(n-1)}2\pi \exp(-x)L^1_{(n-1)}(2x)  {\rm if}\,\, x> 0 ,\,\,
	f_{n}(x) =		0  \,\,\,if \,\,\, x\le 0, 
	\end{equation}
	where $ L^1_{(n-1)}(2x)$ is a Laguerre polynomial.
	Now we see that the functions are equibounded and derivatives of these functions will grow with the growth of $ {n}.  $
	Thus, we have built an example of a sequence of the bounded functions of $ W_{2}^{1}(R)$ which have a constant norm  $ W_{2}^{1}(R)$, and this sequence converges to a discontinuous function.\\
	The results show the flaws of the embedding theorems when analyzing the behavior of functions. Therefore, this work is devoted to overcoming them and the basis for solving the formulated problem is the analytical properties of the Fourier transforms of functions on compact sets. Analytical properties and estimates of the Fourier transform of functions are studied using the Poincaré – Riemann – Hilbert boundary value problem
	
	%------------------------------------------------
	\section{Results for the three-dimensional case }
	Consider Schr\"odinger's equation:
	\begin{equation}
	\label {eq:1}
	-\Delta_{x}\Psi  +q\Psi  =k^{2}\Psi, ~k\in C.  
	\end{equation}
	Let $ \Psi_{+ } (k,\theta,x)  $ be a solution of (\ref{eq:1}) with the following asymptotic behaviour:
	\begin{equation}\label{eq:2}
	\Psi_{+ } (k,\theta,x)=\Psi _{0}(k,    
	\theta, x)+\frac{e^{ ik|x|}}{|x|}A(k,\theta^{'},\theta)+0\left( \frac{1}{|x|}\right), \; |x|\rightarrow \infty,
	\end{equation}
	where   $ A(k,\theta^{'},\theta)  $ is the scattering amplitude and  $ \theta^{'}= \frac{x}{|x|},\;\theta\in S^{2} $ for  $ k\in \bar{C}^{+}=\{{\rm Im} k\ge 0\} $ $ \Psi _{0}(k,\theta, x) =e^{ik(\theta, x)}  $:
	$$\label {eq:3}
	A(k,\theta^{'},\theta)=-\frac{1}{4\pi }\int_{R^{3}}q(x)\Psi _{+ } (k,\theta,x)e^{-ik\theta^{'} x}dx.
	$$
	Solutions to (\ref{eq:1}) and (\ref{eq:2}) are obtained by solving the integral equation
	$$\label{eq:4}
	\Psi_{+ } (k,\theta,x)=\Psi _{0}(k,    
	\theta, x)+\int_{R^3}q(y)\frac{e^{+ ik|x-y|}}{|x-y|}\Psi{+ } (k,\theta,y)dy=G(q\Psi_{+ }), 
	$$
	which is called the Lippman--Schwinger equation.
	
	Let us introduce
	$$ 
	\theta,\theta^{'}\in S^{2}, 
	Df=k\int_{S^{2}}A(k,\theta^{'},\theta)f(k,\theta^{'}) d\theta^{'}.
	$$

	% % % % % % % % % % % % % % % % % % % % % % % % % % % % % % % % % % % % % % % % % % % % % % % % % % % % % % % % % % % % % % % % % %
	Let us also define the solution $ \Psi _{- } (k,\theta,x)$ for $ k\in \bar{C}^{-}=\{{\rm Im} k\le 0\} $ as  \[\Psi _{- } (k,\theta,x)=\Psi _{+ } (-k,-\theta,x) .\]
	As is well known [8], %\cite{a1}
	\begin{equation}
	\label{eq:eq5}
	\Psi _{+ } (k,\theta,x)-\Psi _{- } (k,\theta,x)=
	-\frac{k}{4\pi}\int_{S^{2}}A(k,\theta^{'},\theta)\Psi_{-} (k,\theta^{'},x)d\theta^{'},~  k\in R.
	\end{equation}
	This equation is the key to solving the inverse scattering problem and was first used by Newton~[8,9] and Somersalo et al.~[10].
	
	\begin{definition}
		The set of measurable functions $\mathbf{R}$ with the norm defined by $$
		||q||_{\mathbf{R}}=\sqrt{\int_{R^{6}}\frac{q(x)q(y)}{|x-y|^{2}}dxdy}<\infty $$
		is recognised as being of Rollnik class.
		
	\end{definition}\label{df:d1}

	Equation (\ref{eq:eq5}) is equivalent to the following:
	$$
	\Psi _{+}=S\Psi _{-},  \label{eq:eq6}
	$$
	where  $ S $ is a scattering operator with the kernel 
	$$ ~S(k,\textit{\l})=\int_{R^{3}}\Psi _{+}(k,x)\Psi _{-}^{\ast }(\textit{\l},x)dx. $$
	
	The following theorem was stated in  [9]:
	
	\begin{theorem}\label{thm:t1}\textbf{(Energy and momentum conservation laws)}
	Let $q\in \mathbf{R}$. Then, $ SS^{\ast }=I$ and $S^{\ast }S=I,$ where $ I $ is a unitary operator.
	\end{theorem}
	\begin{corollary}  $ SS^{\ast }=I$ and $S^{\ast }S=I$  yield
	$$
	A(k,\theta^{'},\theta)-A(k,\theta,\theta^{'})^{\ast }=\frac{ik}{2\pi}\int_{S^{2}}A(k,\theta,\theta^{''})A (k,\theta^{'}, \theta^{''} )^{\ast }d\theta^{''}.
	$$
	
	\end{corollary}
	
	\begin{theorem}
	\label{Theorem 1.4}
	\textbf{(Birmann--Schwinger estimation)}
	Let $q\in \mathbf{R}$. Then, the number of discrete eigenvalues can be estimated as 
	$$
	N(q)\leq \frac{1}{(4\pi )^{2}}\int_{R^{3}}\int_{R^{3}}\frac{%
		q(x)q(y)}{|x-y|^{2}}dxdy.
	$$
	\end{theorem}
	%%%
	\begin{lemma}\label{lm:l2} %\Delta=\prod\limits_{j=1}^{N} (k+iE_{j})/(k-iE_{j}
	Let $
	\left(|q|_{L_{1}(R^3)} +4\pi|q|_{L_{2}(R^3)}   \right)<\alpha<1/2   $. Then, 
	$$
	\left\| \Psi_{+}\right\|_{L_{\infty}} \le \frac{\left(|q|_{L_{1}(R^3)} +4\pi|q|_{L_{2}(R^3)}   \right)}{ 1-\left(|q|_{L_{1}(R^3)} +4\pi|q|_{L_{2}(R^3)}   \right)    } < \frac{\alpha}{1-\alpha}, 
	$$
	$$
	\left\|   \frac{ \partial(\Psi_{+}-\Psi_0)}{\partial k}\right\|_{L_{\infty}}\le  \frac{  	|q|_{L_{1}(R^3)} +4\pi|q|_{L_{2}(R^3)}  }{1-\left(|q|_{L_{1}(R^3)} +4\pi|q|_{L_{2}(R^3)}   \right)    }  < \frac{\alpha}{1-\alpha}     
	.$$
	\end{lemma}
	\begin{proof}
	By the Lippman--Schwinger equation, we have
	$$
	\left| \Psi_{+}-\Psi_0\right| \le  \left| Gq\Psi_{+}\right|, \,\,\,\,
	$$
	
	$$
	\left| \Psi_{+}-\Psi_0\right|_{L_{\infty}} \le \left| \Psi_{+}-\Psi_0\right|_{L_{\infty}} \left|   Gq\right|+\left|Gq\right| 	,$$
	and, finally,	
	$$
	\left| \Psi_{+}-\Psi_0\right| \le \frac{\left(|q|_{L_{1}(R^3)} +4\pi|q|_{L_{2}(R^3)}   \right)}{ 1-\left(|q|_{L_{1}(R^3)} +4\pi|q|_{L_{2}(R^3)}   \right)    }.  
	$$	
	
	By the Lippman--Schwinger equation, we also have
	$$
	\left| \frac{ \partial\left( \Psi_{+}-\Psi_0\right) }{\partial k} \right| \le  \left|\frac{ \partial Gq}{\partial k}   \Psi_{+}\right|+  \left| Gq \frac{ \partial\left( \Psi_{+}-\Psi_0\right) }{\partial k}   \right| +\left| Gq\right|,
	$$
	$$\label{eq:eq2}	\left| \frac{ \partial (\Psi_{+}-\Psi_0)}{\partial k} \right|  \le \left(|q|_{L_{1}(R^3)} +4\pi|q|_{L_{2}(R^3)}   \right),  
	$$
	
	$$
	\left\|   \frac{ \partial(\Psi_{+}-\Psi_0)}{\partial k}\right\|_{L_{\infty}}\le  \frac{  	|q|_{L_{1}(R^3)} +4\pi|q|_{L_{2}(R^3)}  }{1-\left(|q|_{L_{1}(R^3)} +4\pi|q|_{L_{2}(R^3)}   \right)    },    
	$$
	which completes the proof.
	\end{proof}
	Let us introduce the following notation:
	$$ 
	Q(k,\theta,\theta^{'})=\int_{R^3}q(x)e^{ik(\theta-\theta^{'})x}dx,\;K(s)=s,\;X(x)=x,\,\,\, 	$$
	$$\label{eq:eq2} T_+Q=\int_{-\infty}^{+\infty}\frac{Q(s,\theta,\theta^{'})}{ s-t-i0}ds,\,\,\,T_-Q=\int_{-\infty}^{+\infty}\frac{Q(s,\theta,\theta^{'})}{ s-t+i0}ds .
	$$

	\begin{lemma}\label{lm:l2} %\Delta=\prod\limits_{j=1}^{N} (k+iE_{j})/(k-iE_{j}
	Let $q\in \mathbf{R}\cap  L_{1}(R^3),\; \left\| q \right\|_{L_1}+4\pi|q|_{L_{2}(R^3)} <\alpha<1/2   $. Then, 
	$$
	\left\|  	A _{+}\right\|_{L_{\infty}}  < \alpha+  \frac{\alpha}{1-\alpha} ,  
	$$
	
	$$
	\left\|  \frac{ \partial A _{+}}{\partial k}	\right\|_{L_{\infty}}  < \alpha+  \frac{\alpha}{1-\alpha}.
	$$
	
	\end{lemma}
	\begin{proof}
	Multiplying the Lippman--Schwinger equation by $q(x)\Psi_0(k,\theta,x)$ and then integrating, we have
	$$
	A(k,\theta,\theta^{'})=	Q(k,\theta,\theta^{'})+		\int_{R^{3}}q(x)\Psi_0(k,\theta,x)G	q\Psi_{+}dx.
	$$
	We can estimate this latest equation as
	$$
	\left| A\right| \le \alpha+ \alpha\frac{\left(|q|_{L_{1}(R^3)} +4\pi|q|_{L_{2}(R^3)}   \right)}{ 1-\left(|q|_{L_{1}(R^3)} +4\pi|q|_{L_{2}(R^3)}   \right)    }.
	$$
	Following a similar procedure for $	\left\|  \frac{ \partial A _{+}}{\partial k}	\right\|$ completes the	proof.	
	\end{proof}
	
	We define the operators ~$T_{\pm }$, $T$ for ~$f\in W_{2}^{1}(R)$ as follows:
	$$ 
	T_{+}f=\frac{1}{2\pi i}\lim\limits_{{\rm Im}z\rightarrow 0}\int\limits_{-\infty}^{\infty }\frac{f({s})}{s-z}ds,~{\rm Im}~z>0,~
	T_{-}f=\frac{1}{2\pi i}\lim\limits_{{\rm Im}z\rightarrow 0}\int\limits_{-\infty
	}^{\infty }\frac{f({s})}{s-z}ds,~{\rm Im}~z<0,
	$$
	$$
	Tf=\frac{1}{2}(T_{+}+T_{-})f.
	$$
	Consider the Riemann problem of finding a function $\Phi $ that is analytic in the complex plane with a cut along the real axis. Values of
	$ \Phi $ on the two sides of the cut are denoted as $\Phi_{+}$ and $\Phi_{-} $. The following presents the results of [12]:
	\begin{lemma}\label{lm:l1}
	$$
	TT= \frac14 I,~TT_+ = \frac12 T_+,~TT_- = - \frac12 T_-, \ T_+ = T+\frac12 I,~T_- = T-\frac12 I,~T_-T_- = -  T_-
	.$$ 
	\end{lemma}
	
	Denote
	$$
	\Phi_{+}(k,\theta,x)=\Psi_{+}(k,\theta,x)-\Psi_{0}(k,\theta,x), \,\,\,\Phi_{-}(k,\theta,x)=\Psi_{-}(k,-\theta,x)-\Psi_{0}(k,\theta,x),\,\, 	$$
	$$\label{eq:eq2} g(k,\theta,x)=\Phi_{+}(k,\theta,x)-\Phi_{-}(k,\theta,x)/
	$$
	
	%\Phi_{-}(k,\theta,x)=\Psi_{-}(k,-\theta,x)-\ph0(k,\theta,x)
	
	\begin{lemma}\label{lm:l2} %\Delta=\prod\limits_{j=1}^{N} (k+iE_{j})/(k-iE_{j}
	Let $q\in \mathbf{R},\,\, N(q)<1, ~g_{+}=g(k,\theta,x)$, and $g_{-}=g(k,-\theta,x). $ Then, 
	$$
	\Phi _{+}(k,\theta,x)=T_{+ } g_{+}  +e^{ik\theta x},\ \Phi _{-}(k,\theta,x)=T_{- } g_{+}  +e^{ik\theta x}.
	$$ 
	\end{lemma}	
	\begin{proof}  
	The proof of the above follows from the classic results for the Riemann problem.
	\end{proof}
	\begin{lemma}\label{lm:l2} %\Delta=\prod\limits_{j=1}^{N} (k+iE_{j})/(k-iE_{j}
	Let $q\in \mathbf{R},\,\, N(q)<1, ~g_{+}=g(k,\theta,x),$ and $g_{-}=g(k,-\theta,x) , ) $. Then, 
	$$
	\Psi _{+}(k,\theta,x)=(T_{+ } g_{+}  +e^{ik\theta x}),\ \Psi _{-}(k,\theta,x)=(T_{- } g_{-}  +e^{-ik\theta x}).
	$$ 
	\end{lemma}
	\begin{proof}
	The proof of the above follows from the definitions of $ g,$ $\Phi _{\pm }$, and $\Psi _{\pm } $ .
	\end{proof}
	%%%%%%%%%%%%%%%%%%%%%%%%%%%%%%%%%%%%%%%
	
	\begin{lemma}\label{lm:l6}
	Let $$ \sup\limits_{ k} \left|    \int\limits_{-\infty}^{\infty} \frac{ pA(p,\theta^{'},\theta) }{ 4\pi(p-k+i0 ) }dp  \right|<\alpha,\,\,\int_{S_2}\alpha d\theta<1/2	.$$ %$$	 \sup\limits_{ k} \left|    \int\limits_{-\infty}^{\infty} \frac{ pA(p,\theta^{'},\theta) \Psi_{0}}{ 4\pi(p-k+i0 ) }dp  \right|<\alpha<1/2 
	%	$$ 
	Then,
	$$
	\prod\limits_{ 0\le j<n }  \int_{S_2}\left|   \int_{-\infty}^{\infty} \frac{ {k_{j}}A(k_{j},\theta^{'}_{k_{j}},\theta_{k_{j}}) }{ 4\pi( k_{j+1}-k_{j}+i0 ) }d{k_{j}} \right |d\theta_{k_{j}} \le 2^{-n}.
	\label{lm:psi}
	$$
	\end{lemma}
	\begin{proof}
	
	Denote 
	$$
	\alpha_j{}=\left| Vp \int_{-\infty}^{\infty} \frac{ {k_{j}}A(k_{j},\theta^{'}_{k_{j}},\theta_{k_{j}}) }{ 4\pi( k_{j+1}-k_{j}+i0 ) }d{k_{j}}\right| , \,\,\, 
	%\beta_{j} = \frac{1}{2}  {k_{j}}A(k_{j},\theta^{'}_{k_{j}},\theta_{k_{j}}).
	$$
	Therefore,
	$$
	\prod\limits_{ 0\le j<n } \int_{S_2} \left| \int_{-\infty}^{\infty} \frac{ {k_{j}}A(k_{j},\theta^{'}_{k_{j}},\theta_{k_{j}}) }{ 4\pi( k_{j+1}-k_{j}+i0 ) }d{k_{j}}  \right |d\theta_{k_{j}}\le   \prod\limits_{ 0\le j<n } \int_{S_2} \alpha_j{} d\theta_{k_{j}}  <
	2^{-n}.	$$
	$$\label{eq:eq2}
	%   as\,\,\, m\neq l ,\alpha_{j_{m}}\neq \alpha_{j_{l}} \,\,\,and \,\,\beta_{j_{m}}\neq \beta_{j_{l}}
	$$
	This completes the proof.
	\end{proof}
	%%%%%%%%%%%%%%%%%%%%%%%%%%%%%%%%%%%%%%%%%%%%%%%%%%%%%%%%%%%%%%%%%%%

	%%%%%%%%%%%%%%%%%%%%%%%%%%%%%%%%%%%%%%%

	\begin{lemma}\label{lm:TA}
	
	Let 	$$
	\sup\limits_{ k } \int_{S^{2}} \left| T_{-}QK\right|d\theta \le \alpha<\frac{1}{2C}<1,\,\,\,\, \sup\limits_{ k } \int_{S^{2}}\left| T_{-}\tilde{q}K\right|d\theta\le \alpha<\frac{1}{2C}<1	,\,\,\,\, 	$$
	$$\label{eq:eq2}\sup\limits_{ k } \int_{S^{2}}\left| T_{-}Q\tilde{q}K^2\right|d\theta\le \alpha<\frac{1}{2C}<1.		$$
	Then,
	$$
	\sup\limits_{ k}\int_{S^{2}}\left| 	T_{-}AK\right|d\theta \le\frac{ C\int_{S^{2}}\left|T_{-}QK\right|d\theta}{1-\sup\limits_{ k}\int_{S^{2}}\left| 	T_{-}A\tilde{q}K^2\right|d\theta},\,\,\, 		$$
	$$\label{eq:eq2}	
	\sup\limits_{ k}\left| \int_{S^{2}}	T_{-}A\tilde{q}K^2d\theta\right| \le
	\frac{ C\left|T_{-}\int_{S^{2}}Q\tilde{q}K^2d\theta\right|}{1- \left| T_{-}\int_{S^{2}}\tilde{q}Kd\theta\right|	}.
	$$

	\end{lemma}
	\begin{proof}
	By the definition of the amplitude and Lemma 4, we have 
	$$
	A(k,\theta^{'},\theta)=-\frac{1}{4\pi }\int_{R^{3}}q(x)\Psi _{+ } (k,\theta,x)e^{-ik\theta^{'} x}dx 	$$
	$$\label{eq:eq2}=-\frac{1}{4\pi }\int_{R^{3}}q(x) \left[ e^{ik\theta^{'} x} +T_+g(k,\theta,\theta^{'}) \right] e^{-ik\theta^{'} x}dx   .
	$$%
	We can rewrite this as
	\begin{equation}\label{eq:5}
	A(k,\theta^{'},\theta)=-\frac{1}{4\pi }\int_{R^{3}}q(x) \left[ e^{ik\theta x} +\sum_{n\ge 0}(-T_-D)^n\Psi_0\right]   e^{-ik\theta^{'} x} dx.
	\end{equation}
	Lemma 6 yields 
	$$
	\sup\limits_{ k}\int_{S^{2}}\left| 	T_{-}AK\right|d\theta\le	\sup\limits_{ k}	\int_{S^{2}}\left| \frac{1}{4\pi }T_{-}QK\right|d\theta  +\frac{\left( \sup\limits_{ k}\int_{S^{2}}\left| 	T_{-}KA\right|d\theta\right) ^2\int_{S^{2}}\left|T_{-} A\tilde{q}K^{2}\right|d\theta }{\left( 1-\sup\limits_{ k}\int_{S^{2}}\left| 	T_{-}KA\right|d\theta\right) ^{2}}. 
	$$
	
	Owing to the smallness of the terms on the right-hand side, the following estimate follows:
	$$
	\sup\limits_{ k}\int_{S^{2}}\left| 	T_{-}AK\right|d\theta\le	2\sup\limits_{ k}	\int_{S^{2}}\left| \frac{1}{4\pi }T_{-}QK\right|d\theta . 
	$$
	Similarly,
	$$\label{eq:eq2}		\sup\limits_{ k}\int_{S^{2}}\left| 	T_{-}A\tilde{q}K^2\right|d\theta \le C\int_{S^{2}}\left|T_{-}Q\tilde{q}K^2\right|d\theta+  \int_{S^{2}}\left| T_{-}A\tilde{q}K^2\right|d\theta	\int_{S^{2}} \left| T_{-}\tilde{q}K\right|d\theta,	   
	$$	
	$$\label{eq:eq2}	
	\sup\limits_{ k}\int_{S^{2}}\left| 	T_{-}A\tilde{q}K^2\right|d\theta \le
	\frac{ C\int_{S^{2}}\left|T_{-}Q\tilde{q}K^2\right|d\theta}{1- \int_{S^{2}}\left| T_{-}\tilde{q}K\right|d\theta	},
	$$
	$$
	\sup\limits_{ k}\int_{S^{2}}\left| 	T_{-}A\tilde{q}K^2\right|d\theta\le	2\sup\limits_{ k}\int_{S^{2}}	\left| \frac{1}{4\pi }T_{-}Q\tilde{q}K^2\right|d\theta  .
	$$

	This completes the proof.	
	\end{proof}
	%%%%%%%%%%%%%%%%%%%%%%%%%%%%%%%%%%%%%%%%%%%%%%%%%%%%%%%%%%%%%%%%%%%%%
	
	%%%%%%%%%%%%%%%%%%%%%%%%%%%%%%%%%%%%%%%%

	%%%%%%%%%%%%%%%%%%%%%%%%%%%%%%%%%%%%%%%%%%%%%%%%%%%%%%%%%%%%%%%%%%%%%%%%%%%%%%%%%%%%%%%%%%%%%%%%%
	To simplify the writing of the following calculations, we introduce the set defined by
	$$M_{\epsilon}(k)=\left( s|  \epsilon<|s|+|k-s|<\frac{1}{\epsilon}\right) .$$
	The Heaviside function is given by 
	$$
	{\Theta}(x) =
	\left\{
	1 ,\,\,\mbox{if } x> 0 ,\,\,\,
	 \,\,\,\,
	-1 \,\,  \mbox{if } x< 0 \,\,\,
	\right\}.
	$$

	\begin{lemma} \label{lm:l8}
	Let $q,\nabla q\in \cap L_{2}(R^{3})$, $|A|>0$. Then,
	$$ 
	\pi i\int_{R^3} \Theta(A) e^{ik|x|A}q(x)dx  =    \lim\limits_{\epsilon\rightarrow 0}\int_{s\in M_{\epsilon}(k)}\int_{R^3} \frac{e^{is|x|A}}{k-s} q(x)dxds,   
	$$
	$$ 
\pi i	\int_{R^3} \Theta(A) ke^{ik|x|A}q(x)dx  =    \lim\limits_{\epsilon\rightarrow 0}\int_{s\in M_{\epsilon}(k)}\int_{R^3} s\frac{e^{is|x|A}}{k-s} q(x)dxds  . 
	$$
	
	\end{lemma} 
	\begin{proof}
	The lemma can be proved by the conditions of lemma and the lemma of Jordan.
	\end{proof}

	%%%%%%%%%%%%%%%%%%%%%%%%%%%

	%%%%%%%%%%%%%%%%%%%%%%

	%%%%%%%%%%%%%%%%%%%%%%%%%%%%%%%%%%%%%%%%%%%%%%%%%%%%%%
	\begin{lemma}\label{lm:TkQ}
	
	Let $$ \,\,\,I_0= \Psi_0(x,k)|_{ r=r_{0}} .
	$$
	Then
	$$		\left| 	\int_{-\infty}^{+\infty}\int_{S^{2}}\int_{S^{2}}\tilde{q}(k(\theta-\theta'))I_0k^2dkd\theta d\theta'\right| \le \sup\limits_{x\in R^3} \left| q(x)\right| +C_0(\frac{1}{r_{0}} +r_0)\left\| q\right\|_{L_2(R^3)}, 
	$$
	
	$$	 	\sup \limits_{\theta\in S^2}\left|\int_{-\infty}^{+\infty} \int_{S^{2}}\int_{S^{2}} QTKQI_0k^2d\theta''d\theta'dk \right|  \le  C_0(\frac{1}{r_{0}} +r_0)\left\| q\right\|^2_{L_2(R^3)} .
	$$	
	\end{lemma}
	\begin{proof}
	By the definition of the Fourier transform, we have
	$$\label{eq:eq2}	
	\int_{-\infty}^{+\infty}\int_{S^{2}}\int_{S^{2}}\tilde{q}(k(\theta-\theta'))I_0k^2dkd\theta d\theta'=
	\int_{-\infty}^{+\infty}\int_{S^{2}}\int_{S^{2}} \int_{0}^{+\infty}q(x)e^{ikx(\theta-\theta')}e^{ix_0k}k^2dkd\theta d\theta' drd\gamma,
	$$	
	where $x=r\gamma$
	The lemma of Jordan completes the proof for the first inequality.
	The second inequality is proved like the first:
	$$	Vp \int_{-\infty}^{+\infty} \int_{S^{2}}\int_{S^{2}} QTKQI_0k^2d\theta''d\theta'dk 
	$$
	$$
	=Vp\int_{-\infty}^{+\infty} \int_{-\infty}^{+\infty} \int_{S^{2}}\int_{S^{2}}\int_{S^{2}} \frac{\left( \tilde{q}(s\cos(\theta')- s\cos(\theta''))\tilde{q}(k\cos(\theta)-s\cos(\theta'')\right)s }{k-s}   I_0k^2d\theta'd\theta''d\theta dk ds.
	$$				
	Lemma 8 yields			
	$$	\int_{-\infty}^{+\infty}\int_{S^{2}} \int_{S^{2}}\int_{S^{2}} \left( \tilde{q}(k\cos(\theta')- k\cos(\theta))\tilde{q}(k\cos(\theta)-k\cos(\theta'')\right) I_0k^3\Theta (\cos(\theta''))d\theta'd\theta''d\theta dk-
	$$
	$$	\int_{-\infty}^{+\infty} \int_{S^{2}}\int_{S^{2}} \int_{S^{2}}\left( \tilde{q}(k\cos(\theta')- k\cos(\theta))\tilde{q}(k\cos(\theta)-k\cos(\theta'')\right) I_0k^3\Theta (-\cos(\theta''))d\theta'd\theta''d\theta dk.
	$$
	Integrating $\theta$, $\theta'$, $\theta''$, and $k$, we obtain the proof of the second inequality of the lemma.
	
\end{proof}

%%%%%%%%%%%%%%%%%%%%%%%%%%%%%%%%%%%%%%

\begin{lemma}\label{lm:TA}
	
	Let 	$$
	\sup\limits_{ k }  \left| T_{-}QK\right| \le \alpha<\frac{1}{2C}<1,\,\,\,\, \sup\limits_{ k } \left| T_{-}\tilde{q}K\right|\le \alpha<\frac{1}{2C}<1	,\,\,\,\, 	$$
	$$\label{eq:eq2}\sup\limits_{ k } \left| T_{-}Q\tilde{q}K^2\right|\le \alpha<\frac{1}{2C}<1	,\,\,\,l=0,1,2	.	$$
	Then,
	
	$$\label{eq:A^0}	
	\left| 	\int_{-\infty}^{+\infty} \int_{S^{2}}\int_{S^{2}} A(k,\theta',\theta)k^ldkd\theta'd\theta \right| \le \left| \int_{-\infty}^{+\infty}\int_{S^{2}}\int_{S^{2}}\tilde{q}(k(\theta-\theta'))k^ldkd\theta' d\theta\right| 
	$$$$
	+C\sup \limits_{\theta\in S^2}\left|\int_{-\infty}^{+\infty} \int_{S^{2}}\int_{S^{2}} QTKAk^ld\theta''d\theta'dk \right|, 
	$$
	$$\label{eq:A^0}	
	\left| 	\int_{-\infty}^{+\infty} \int_{S^{2}}\int_{S^{2}} A(k,\theta',\theta)k^2dkd\theta'd\theta \right| \le \sup\limits_{x\in R^3} \left| q\right|  + C_0\left\| q\right\|_{W_2^1(R^3)}\left\| q\right\|_{L_2(R^3)}\left(  \left|\int_{S^{2}}TKAd\theta'' \right| +1  \right) 
	.$$

\end{lemma}
\begin{proof}

	Using the definition of the amplitude, Lemmas 3 and 4, and the lemma of Jordan yields 
	$$
	\int_{-\infty}^{+\infty}\int_{S^{2}}\int_{S^{2}}	A(k,\theta^{'},\theta)k^ldkd\theta'd\theta =-	\int_{-\infty}^{+\infty}\frac{1}{4\pi }\int_{S^{2}}\int_{S^{2}}\int_{R^{3}}q(x)\Psi _{+ } (k,\theta,x)e^{-ik\theta^{'} x}k^ldxdkd\theta' =	$$
	$$\label{eq:eq2}-\frac{1}{4\pi }\int_{S^{2}}\int_{S^{2}}\int_{R^{3}}q(x) \left[ e^{ik\theta x} +\sum_{n\ge  1}(-T_-D)^n\Psi_0\right]  e^{-ik\theta^{'} x}k^ld\theta' dxdk 
	$$$$
	=\int_{-\infty}^{+\infty}\int_{S^{2}}\int_{S^{2}}\tilde{q}(k(\theta-\theta'))k^ldkd\theta'd\theta+ \sum_{n\ge  1}W_n 
	,$$%	
	$$
	W_1= Vp\int_{R^{3}}\int_{-\infty}^{+\infty}\int_{S^{2}}\int_{S^{2}}\frac{sA(s,\theta^{''},\theta)e^{-ik\theta^{'} x}q(x)e^{i s\theta'' x }}{k-s}k^l dkdxdsd\theta'd\theta'',
	$$
	
	$$
	\left| 	W_1\right| \le  C\sup \limits_{\theta\in S^2}\left|\int_{-\infty}^{+\infty} \int_{S^{2}}\int_{S^{2}} QTKAk^ld\theta''d\theta'dk \right|  
	.$$

	Similarly, 
	$$
	\left| 	W_n\right| \le C\sup \limits_{\theta\in S^2}\left|\int_{-\infty}^{+\infty} \int_{S^{2}}\int_{S^{2}} QTKAk^ld\theta''d\theta'dk \right| \left|\int_{S^{2}}TKAd\theta'' \right|^n
	.$$
	Finally,
	
	$$
	\left| 	\int_{-\infty}^{+\infty} \int_{S^{2}}\int_{S^{2}} A(k,\theta',\theta)dkd\theta'd \theta \right| \le \left| \int_{-\infty}^{+\infty}\int_{S^{2}}\int_{S^{2}}\tilde{q}(k(\theta-\theta'))dkd\theta d\theta'\right|
	$$$$
	+C_0\left\| q\right\|^2_{L_2(R^3)}\left(  \left|\int_{S^{2}}TKAd\theta'' \right| +1  \right), 
	$$
	
	$$
	\left| 	\int_{-\infty}^{+\infty} \int_{S^{2}}\int_{S^{2}} A(k,\theta',\theta)k^2dkd\theta'\right| \le \sup\limits_{x\in R^3} \left| q\right|  + C_0\left\| q\right\|^2_{L_2(R^3)}\left(  \left|\int_{S^{2}}TKAd\theta'' \right| +1  \right). 
	$$
	%%%%%%%%%%%%%%%%%%%%%%%%%%%%%%%%%%%%5

	This completes the proof.	
\end{proof}

\begin{lemma}\label{lm:l6}
	Let $$  \sup\limits_{ k}  \int_{S^{2}}\left|    \int\limits_{-\infty}^{\infty} \frac{pA(p,\theta^{'},\theta) }{ 4\pi(p-k+i0 ) }dp  \right|d\theta<\alpha<1/2,\,\,\, \sup\limits_{ k} \left| pA(p,\theta^{'},\theta)\right|<\alpha<1/2. 
	$$  Then,
	$$
	|T_{- }D\Psi _{0} |< \frac{\alpha }{1-\alpha},\,\,\,\, |T_{+ }D\Psi _{0} |< \frac{\alpha }{1-\alpha}, \,\,\,\,\,\,\,\,\,
	|D\Psi _{0} |< \frac{\alpha }{1-\alpha}, 	$$
	$$\label{eq:eq2}
	\,\, 	T_{- }g_{-}=( I-T_{- }D )^{-1 }T_{- }D\Psi _{0},
	\,\,\,\,\,\,\,\,\,\,\,\,\,\Psi _{-}=( I-T_{- }D )^{-1 }T_{- }D\Psi _{0} +\Psi _{0},           
	\label{lm:psi}
	$$
	and $q$ satisfies the following inequalities:
	$$
	\sup\limits_{x\in R^3}|q(x)|\le \left| \int_{S^{2}}TKQ d\theta\right|
	C_0\left( \left\| q\right\|^2_{L_2(R^3)}+1\right) +C_0\left\| q\right\|_{L_2(R^3)}.
	$$
\end{lemma}
\begin{proof}
	Using the equation 
	$$
	\Psi _{+ } (k,\theta,x)-\Psi _{- } (k,\theta,x)=
	%$$
	%$$
	-\frac{k}{4\pi}\int_{S^{2}}A(k,\theta^{'},\theta)\Psi_{-} (k,\theta^{'},x)d\theta^{'},~  k\in R,
	$$%
	we can write 
	$$
	T_{+}g_{+}-T_{- }g_{-}=D (T_{-}g_{-}+\Psi_{0}).
	$$
	Applying the operator $T_{-}$ to the last   equation, we have 
	$$
	T_{- }g_{-}=T_{- }D (T_{-}g_{-}+\Psi_{0}),
	$$
	$$
	(I  - T_{- }D    ) T_{- }g_{-}=T_{- }D\Psi_{0},\,\,\,
	T_{- }g_{-}=   \sum_{n\ge0}      \left( -T_{- }D \right)^{n} \Psi_{0}.
	$$
	Estimating the terms of the series, we obtain using Lemma 4
	$$
	|(T_{- }D)^{n}\Psi_{0}| \le 
	\sum_{n\ge 0 }
	\left|  \int_{-\infty}^{\infty}\dots \int_{-\infty}^{\infty} \Psi_{0}  \prod\limits_{ 0\le j<n }\frac{ \int_{S^{2}}{k_{j}}A(k_{j},\theta^{'}_{k_{j}},\theta_{k_{j}}) d\theta^{'}_{k_{j}}}{ 4\pi(k_{j+1})-k_{j}+i0 ) }   dk_{1}\dots d_{k_{n}}\right| 	$$
	$$\label{eq:eq2}\\\le \sum_{n> 0 } 2^n \alpha^{n} =\frac{2\alpha}{1-2\alpha} . 
	$$
	%	Denote $$\Lambda= \frac{1}{k^2}  \frac{\partial }{\partial k}k^2\frac{\partial }{\partial k}, r=\sqrt{x_1^2+x_2^2+x_3^2}, $$ 
	Denoting $$\Lambda=  \frac{\partial }{\partial k},\; r=\sqrt{x_1^2+x_2^2+x_3^2}, $$ 
	we have
	$$
	\Lambda\int_{S^{2}}\Psi_0d\theta =\Lambda  \frac{\sin( kr)}{ikr}=  
	\frac{\cos(kr)}{ik}-\frac{\sin( kr)}{ik^2r}, 
	$$
	$$
	\Lambda\int_{S^{2}}H_0\Psi_0d\theta =\Lambda k^2 \frac{\sin( kr)}{ikr}=  
	k\frac{\cos(kr)}{i}+\frac{\sin( kr)}{ik^2r} ,
	$$
	$$
	\left| \Lambda \int_{S^{2}}\Psi d\theta	 \right|=\left|\Lambda\int_{S^{2}}\Psi_0d\theta +\Lambda\int_{S^{2}}\sum_{n\ge 0}\left( -T_{- }D \right)^{n} \Psi_{0} d\theta\right|  > \left(\frac{1}{k} -  \frac{\alpha}{1-\alpha}\right), \;{\rm as}\,\, kr=\pi,
	$$
	and
	$$
	\Lambda\frac{1}{k-t}=
	-   \frac{1}{(k-t)^2} 
	$$

	Equation (\ref{eq:1})  yields 
	$$
	q= \frac{\Lambda\left(  H_0\int_{S^{2}}\Psi d\theta+k^2\int_{S^{2}}\Psi d\theta\right)  }{\Lambda\int_{S^{2}}\Psi d\theta} $$$$
	=\frac{2k\int_{S^{2}}T_{-}g_{-}d\theta+k^2\int_{S^{2}}\Lambda T_{-}g_{-}d\theta+  H_0\Lambda \int_{S^{2}}T_{-}g_{-}d\theta}{\Lambda\int_{S^{2}}\Psi d\theta}  $$
	$$=
	\frac{2k\int_{S^{2}}T_{-}g_{-}d\theta+\Lambda\int_{S^{2}}\sum_{n\ge 1}      \left( -T_{- }D \right)^{n}(K^2-k^2) \Psi_{0} d\theta}{\Lambda\int_{S^{2}}\Psi d\theta}$$$$
	=\frac{W_0+\sum_{n\ge 1} \int_{S^{2}}W_{n}}{\Lambda\int_{S^{2}}\Psi d\theta}.
	$$

	Denoting $$Z(k,s)=s+2k +\frac{2k^2}{k-s},$$ we then have 
	$$
	\left|  W_1\right| \le 	\left|Vp\int_{-\infty}^{+\infty}\int_{S^{2}} \int_{S^{2}} A(s,\theta,\theta')s \frac{s^2-k^2}{(k-s)^2}  \Psi_0  \sin(\theta)ds d\theta\right|_{k=k_0} 
	$$$$
	\le \left| \int_{-\infty}^{+\infty}\int_{S^{2}}\int_{S^{2}}Z(k,)\tilde{q}(k(\theta-\theta'))\Psi_0dkd\theta\right|+
	C_0\left|\int_{S^{2}} TKQd\theta\right|.
	$$
	
	For calculating $W_{n}$, as $n \ge 1$, take the simple transformation
	$$
	\frac{s_{n}^3}{s_{n}-s_{n-1}}=	\frac{s_{n}^3-s_{n}^2s_{n-1}}{s_{n}-s_{n-1}} +\frac{s_{n}^2s_{n-1}}{s_{n}-s_{n-1}}=s_{n}^2+\frac{s_{n}^2s_{n-1}}{s_{n}-s_{n-1}}
	$$
	\begin{equation} \label{eq:6}
	=s_{n}^2+\frac{s_{n}^2s_{n-1}-s_{n}s_{n-1}^2}{s_{n}-s_{n-1}} +\frac{s_{n}s_{n-1}^2}{s_{n}-s_{n-1}}=	s_{n}^2+s_{n}s_{n-1}+\frac{s_{n}s_{n-1}^2}{s_{n}-s_{n-1}},
	\end{equation}
	$$
	\frac{As_{n}^3}{s_{n}-s_{n-1}} =As_{n}^2+As_{n}s_{n-1}+    \frac{As_{n}s^2_{n-1}}{s_{n}-s_{n-1}}=V_1+V_2+V_3 .
	$$
	Using Lemma \ref{lm:TA}   for estimating $V_1$  and $V_2$ and, for $ V_3 $, taking again the simple transformation for $s^3_{n-1} $, which will appear in the integration over  $ s_{n-1}$, we finally get 
	$$
	|q(x)|_{r=r_0}= \left| \frac{\Lambda\left(  H_0\int_{S^{2}}\Psi d\theta+k^2\int_{S^{2}}\Psi d\theta\right)  }{\Lambda\int_{S^{2}}\Psi d\theta}\right| _{k=k_0,r= \frac{\pi}{k_0}} $$$$
	\le \frac{\left| \int_{-\infty}^{+\infty}\int_{S^{2}}\int_{S^{2}}Z(k,)\tilde{q}(k(\theta-\theta'))\Psi_0dkd\theta d\theta'\right|+
		C_0\left|\int_{S^{2}} TKQd\theta\right|}{(\frac{1}{k_0}-\frac{\alpha}{(1-\alpha)} )}   + 
	$$
	$$
	% \frac{   C_0\left\| q\right\|_{W_2^1(R^3)}\left\| q\right\|_{L_2(R^3)}\left(  %\left|\int_{S^{2}}TKAd\theta'' \right| +1  \right)}{1-\frac{\alpha}{r_0^2-\frac{\alpha}{(1-\alpha)k_0}} } 
	$$	
	Finally, we get
	$$
	|q(x)|_{r=r_0}\le \sup\limits_{x\in R^3}|q(x)|\alpha+
	C_0\left\| q\right\|^2_{L_2(R^3)}+C_0\left\| q\right\|_{L_2(R^3)}+\left| \int_{S^{2}}TKQ d\theta\right|.
	$$
	
	The invariance of the Schr\"odinger equations with respect to translations and the arbitrariness of $r_0$	yield
	$$
	\sup\limits_{x\in R^3}|q(x)|\le \left| \int_{S^{2}}TKQ d\theta\right|
	C_0\left( \left\| q\right\|^2_{L_2(R^3)}+1\right) +C_0\left\| q\right\|_{L_2(R^3)}.
	$$
\end{proof}
%%%%%%%%%%%%%%%%%%%%%%%%%%%%%%%%%%%%%%%%%%%%%%%%%	

%%%%%%%%%%%%%%%%%%%%%%%%%%%%%%%%%%%%%%%%%%%%%%%%%%%%%%%%%%%%%%%%%%%%%%%%%%%%%%%%%%%%%%%%%%%%%%%%%

%------------------------------------------------
% % % % % % % % % % % % % % % % % % % % % % % % % % % % % % % % % % % % %
% % % % % % % % % % % % % % % % % % % % % % % % % % % % % % % % % % %5
% % % % % % % % % % % % % % % % % % % % % % % % % % % % % % %
% % % % % % % % % % % % % % % % % % % % % % % % % % %
\section{Discussion of the three-dimensional inverse scattering problem }
This study has shown, once again, the outstanding properties of the scattering operator, which, in combination with the analytical properties of the wave function, allows us to obtain almost-explicit formulas for the potential  from the scattering amplitude. Furthermore, this appro. The estimations following from this overcome the problem of overdetermination, resulting from the fact that the potential is a function of three variables, whereas the amplitude is a function of five variables. We have shown that it is sufficient to average the scattering amplitude to eliminate the two extra variables.
%------------------------------------------------
\section{  Studying the properties of solutions of the Cauchy problem for the Navier--Stokes equations using analytic functions generated by the Schr\"odinger equations and related to the Poincar\'e-–-Riemann–-Hilbert problem}
Numerous studies of the Navier--Stokes equations have been devoted to the problem of the smoothness of its solutions. A good overview of these studies is given in Refs. [13--17]. The spatial differentiability of the solutions is an important factor, as it controls their evolution. Obviously, differentiable solutions do not provide an effective description of turbulence. Nevertheless, the global solvability and differentiability of the solutions have not been proven, and therefore the problem of describing turbulence remains open. It is interesting to study the properties of the Fourier transform of solutions of the Navier--Stokes equations. Of particular interest is how they can be used in the description of turbulence and whether they are differentiable. The differentiability of such Fourier transforms appears to be related to the appearance or disappearance of resonance, as this implies the absence of large energy flows from small to large harmonics, which in turn precludes the appearance of turbulence.
Therefore, obtaining uniform global estimations of the Fourier transform of solutions of the Navier--Stokes equations means that the principle modelling of complex flows and related calculations will be based on the Fourier transform method. We are continuing to research these issues in relation to a numerical weather prediction model; this paper provides a theoretical justification for this approach. 

Consider the Cauchy problem for the Navier--Stokes equations:
\begin{equation}
\label{1}
%q_{t}-\nu \Delta q+\sum\limits_{k=1}^{3}q_{k}q_{x_{k}}=-\nabla
\frac{\partial\vec{v}}{\partial t}-\nu \Delta \vec{v}+(\vec{v},\nabla \vec{v})=-\nabla
p+\vec{f}(x,t),~{\rm div}~\vec{v}=0,  
\end{equation}
\begin{equation}\label{2}
\vec{v}|_{t=0}=\vec{v}_{0}(x)  
\end{equation}
in the domain $Q_{T}=R^{3}\times (0,T)$, where 
\begin{equation}
{\rm div}\;\vec{v}_{0}=0.  \label{3}
\end{equation}
The problem defined by  (\ref{1})--(\ref{3}) has at least one weak solution $ (\vec{v}, p) $ in the so-called Leray--Hopf class [16].
Denote:
$$ \mu(x)=\sqrt{1+|x|} , \,\,\,|x|=\sqrt{\sum_{1}^{3}}x^2_i$$ 
The following results have been  proved  [15]:
\begin{theorem}
	\textbf{} If
	$$
	\mu\vec{v}_{0}\in W_{2}^{1}(R^{3}),\mu\vec{f}(x,t)\in L_{2}(Q_{T}),
	$$%
	there is a single generalised solution of (\ref{1})--(\ref{3}) in the domain $Q_{T_{1}}$, $T_{1}\in \lbrack 0,T]$, satisfying the following conditions:
	$$
	\mu\vec{v},\mu\nabla ^{2}\vec{v},\ \ \ \mu\nabla p\in L_{2}(Q_{T}).
	$$
	\label{thm1}
\end{theorem}
Note that $T_{1}$ depends on $\vec{v}_{0}$ and $\vec{f}(x,t)$.
\begin{lemma}\label{lm:l8}
Let $\mu\vec{v_0} \in W_{2}^{2}(R^{3}),\mu\vec{f} \in L_2(Q_T)  $,
	then the solution of (\ref{1})--(\ref{3}) satisfies  the following inequalities:
	$$
	\sup\limits_{0\leq t\leq
		T}||\mu\vec{v}||_{L_{2}(R^{3})}^{2}+\nu\int\limits_{0}^{t}||\mu\nabla \vec{v}||_{L_{2}(R^{3})}^{2}d\tau  \leq \ ||\mu\vec{v}_{0}||_{L_{2}(R^{3})}^{2}+||\mu\vec{f}||_{L_{2}(Q_{T})},	$$
	$$\label{eq:eq2}
	\sup\limits_{0\leq t\leq
		T}||\mu\vec{\nabla v}||_{L_{2}(R^{3})}^{2}+\nu\int\limits_{0}^{t}||\mu H_{0} \vec{v}||_{L_{2}(R^{3})}^{2}d\tau  	$$
	$$\label{eq:eq2}\leq || \mu\nabla\vec{v}_{0}||_{L_{2}(R^{3})}^{2}+||\mu\vec{f}||_{L_{2}(Q_{T})} +\int_{0}^{t}||\mu(\vec{v},\nabla \vec{v})||_{L_{2}(R^{3})}|| \mu H_{0} \vec{v}||_{L_{2}(R^{3})},	$$
	$$\label{eq:eq2}
	\nu\int\limits_{0}^{t}||\mu H_{0} \vec{v}||_{L_{2}(R^{3})}^{2}d\tau \le C+ \frac{1}{\nu}\int_{0}^{t}||(\mu\vec{v},\nabla \vec{v})||^{2}_{L_{2}(R^{3})}dt	.
	$$
\end{lemma}
\begin{lemma}
	Let $\vec{v_0} \in W_{2}^{2}(R^{3}),$ $\vec{\tilde{v_0}} \in W_{2}^{2}(R^{3}),$ and $\vec{f} \in L_2(Q_T) $.
	Then, the solution of (\ref{1})--(\ref{3}) satisfies the following:
	$$
	\widetilde{\vec{v}}=\widetilde{\vec{v}}_{0}+
	\int\limits_{0}^{t}e^{-\nu k^{2}|(t-\tau )}(\widetilde{[(\vec{v},\nabla )\vec{v}]}+\widetilde{\vec{F}})
	d\tau ,
	\label{eqno8.4}
	$$
	where $\vec{F}=-\nabla p+\vec{f}$.
\end{lemma}
\begin{proof}
	This follows from the definition of the Fourier transform and the theory of linear differential equations.
\end{proof}

% % % % % % % % % % % % % % % % % % % % % % % % % %55
% % % % % % % % % % % % % % % % % % % % % % % % % % %5

Let us introduce the operators $F_{k}$ and $F_{kk\prime}$ as $$F_{k}f=\int_{R^{3}} e^{i(k,x)} f(x)dx,\,\,\,F_{kk\prime}f=\int_{R^{3}} e^{i(k,x)-i(x,k^{\prime})} f(x)dx, $$
$$\vec{\tilde{v}} (k)=F_{k}\vec{v},\,\, \vec{V} (k,k^{\prime})=F_{kk\prime}\vec{v}= \int_{R^{3}} e^{i(k,x)-i(x,k^{\prime})}\vec{v}dx.$$
%%%%%%%%%%%%%%%%%%%%%%%%%%%%%%%%%%%%%%%%%%%%%%%%%%%%%%%%%%%%%%%%%%%%%%%%%%%%%%%%%%%%%%%%%%%%%%%%%\

%%%%%%%%%%%%%%%%%%%%%%%%%%%%%%%%%%%%%%%%%%%%%%%%%%%%%%%%%%%%%%%%%%%%%%%%%%%%%%%%%%%%%%%%%%%%%%%%%\
\begin{lemma}\label{lm:19}
	Let $\vec{v_0} \in W_{2}^{2}(R^{3})$, $\vec{f}\in L_2(Q_T)$, and $ \left| TKV_0\right| +\left| TKV_0\right| +\left| TK^2V_0\vec{\tilde{v_0}}\right| <C$. Then,
	the solution of (\ref{1})--(\ref{3}) in Theorem \ref{thm1} satisfies  the following inequalities:
	$$
	|\tilde{v} (k)|<C,\,\,
	\,\,
	$$
	$$\label{eq:eq2}
	|TK\tilde{v} (k)|<C_0||v||_{L_2(R^3)}+   \frac{C_0t}{\sqrt{\nu}}||\nabla v||_{L_2(R^3)}||v||_{L_2(R^3)}
	.$$
\end{lemma}
\begin{proof}
	This follows from 
	$$
	\vec{\dot v}=
	-(\vec{v}\nabla )\vec{v}+(\nu  \vec{v} + \nabla p) + F ,\\
	$$	
	$$
	\vec{\tilde{v}} = \vec{\tilde{v}}_0+
	\int_{0}^{t} e^{-\nu k^2(t-\tau)}F_{k}\left( -\ (\vec{v},\nabla )\vec{v})+	  \nabla p + 	F \right)d\tau .
	$$	
	From the last equation we have
	$$
	|\vec{\tilde{v}}|= |\le |\vec{\tilde{v}}_0|+C_{T}.
	$$

	Denote $$\beta=\sqrt{\nu(t-\tau)},\,\,\,a=\theta x$$
	formula  121 (23) from [11] as $n=0$: yield	
	$$
	\left| 	TK\vec{v} \right| <\left| k e^{-\beta^2k^2}\right|  + \sqrt{\pi}\beta^{-1}e^{- \frac{a^2}{8\beta^2}}D_{0}\left(\frac{a}{\sqrt{2}\beta }\right), 		$$
	$$\label{eq:eq2}
	\left| TK\vec{v} \right| \le	\left| TK\vec{v}_0\right| 	$$
	$$\label{eq:eq2}	
	+\left| TK\int_{0}^{t} e^{-\nu k^2(t-\tau)}F_{k}\left( -	 (\vec{v},\nabla )\vec{v}]+ \nabla p + 	F\right)dk\right|	$$
	$$\label{eq:eq2}	
	\le \left| TK\vec{v}_0\right| +
	\int_{0}^{t}\left| k e^{-\beta^2k^2}\right|  +	\left|  \sqrt{\pi}\beta^{-1}e^{- \frac{a^2}{8\beta^2}}D_{0}(\frac{a}{\sqrt{2}\beta }) \right|||\nabla \vec{v}||_{L_{2}(R^3)}dt 
	$$$$
	\le C_0||v||_{L_2(R^3)}+   \frac{C_0t}{\sqrt{\nu}}||\nabla v||_{L_2(R^3)}||v||_{L_2(R^3)}	.
	$$

\end{proof}
%%%%%%%%%%%%%%%%%%%%%%%%%%%%%%%%%%%%%%%%%%%%%%%%%%%%%%%%%%%%%%%%%%%%%

\begin{lemma}\label{lm:19_15}
	Let $A>0,p>0,q>0 \,\,\,1/p+1/q=1$
	then
$$	\left| Vp\int\limits_{-\infty}^{\infty}\frac{e^{-Ak^2}}{k-l}dk \right| <\frac{C_p}  {  {A^{1/2p}}  }+C_0    e^{-Al^2} $$
$$	\left| Vp\int\limits_{-\infty}^{\infty}\frac{ke^{-Ak^2}}{k-l}dk \right| <\frac{C_p}  {  {A^{1/p}}  }+C_0    (|l|+1)e^{-Al^2}  $$
	
\end{lemma}
\begin{proof}
	
$$	\left| Vp\int\limits_{-\infty}^{\infty}\frac{e^{-Ak^2}}{k-l}dk \right| <\left| Vp\int\limits_{|k-l|>|1}\frac{e^{-Ak^2}}{k-l}dk \right| + \left| Vp\int\limits_{|k-l|<1}\frac{e^{-Ak^2}}{k-l}dk \right|  <  $$	
$$	\left|  \int\limits_{ |k-l|>1}{e^{-pAk^2}}dk \right|^{1/p}\left|  \int\limits_{ |k-l|>1}\frac{1}{|k-l|^{q}}dk \right|^{1/q} + \left| Vp\int\limits_{|k-l|<1}\frac{e^{-Ak^2}}{k-l}dk \right|   $$

$$\left| Vp\int\limits_{|k-l|<1}\frac{e^{-Ak^2}}{k-l}dk \right| <
\left|\int\limits_{k=l+e^{i\phi}}\frac{e^{-Ak^2}}{k-l}dk \right|
+C_0e^{-Al^2}
  $$	
\end{proof}
	Denote $$\beta=\sqrt{(1-\cos(\theta))(t-\tau)\nu}$$
$$ R=-	F_{kk\prime} (\vec{v},\nabla )\vec{v}]+	 F_{kk\prime} \nabla p + 	F_{kk\prime}{F} $$

\begin{lemma}\label{lm:19}
	Let $\vec{v_0} \in W_{2}^{2}(R^{3})$, $\vec{f}\in L_2(Q_T)$,  $ \left| TKV_0\right| +\left| TK^2V_0\vec{\tilde{v_0}}\right|
	.$ Then,
	the solution of (\ref{1})--(\ref{3}) in Theorem \ref{thm1} satisfies  the following inequalities:
	$$
	|\vec{V}  (k,k^{\prime})|<C,\,\,
	k|\vec{V}  (k,k^{\prime})|<\frac{C}{ \sqrt{(1-\cos(\theta))}}	,\,\,	$$
	$$\label{eq:eq2}	 |T\vec{V} K|<C_0||v||_{L_2(R^3)}+   \int_{0}^{t} \frac{1}{\beta} \left(\int\limits_{R^3}|x||v\nabla v|dx + ||\nabla v||_{L_2(R^3)}||v||_{L_2(R^3)}  \right) d\tau	$$
		$$\label{eq:eq2}	 |T\vec{V} K|<C_0||v||_{L_2(R^3)}+    \frac{t}{\sqrt{(1-\cos(\theta))}}  \max\limits_{ 0\le \tau<t } \left(\int\limits_{R^3}|x||v\nabla v|dx + ||\nabla v||_{L_2(R^3)}||v||_{L_2(R^3)}  \right) 	$$
\end{lemma}
\begin{proof}
	This follows from 
	$$
	\vec{\dot V}=
	-F_{kk\prime}[(\vec{v},\nabla )\vec{v}]+F_{kk\prime} (\nu\Delta \vec{v} +\nabla p) + F_{kk\prime}{F} .
	$$
	
	After the transformations, we obtain
	$$
	\vec{\dot V}=
	-F_{kk\prime}[(\vec{v}\nabla )\vec{v}]+(\nu_{k} F_{kk\prime}  \vec{v} +F_{kk\prime} \nabla p) + F_{kk\prime}{F} ,\\
	$$	
	$$
	\vec{V} = \vec{V}_0+
	\int_{0}^{t} e^{-\nu k^2(1-\cos(\theta))(t-\tau)}\left( -	F_{kk\prime}[(\vec{v},\nabla )\vec{v}]+	 F_{kk\prime} \nabla p + 	F_{kk\prime}{F} \right).
	$$	
	From the last equation, we have
	$$
	|\vec{V} |\le |\vec{V} _0|+C_0\int_{0}^{t}||\nabla v||_{L_2(R^3)}||v||_{L_2(R^3)}	d\tau.
	$$

 Lemma   (\ref{lm:19_15})  yield	

	$$\label{eq:eq2}
	\left| TK\vec{V} \right| \le	\left| TK\vec{V}_0\right|+
	\left| TK\int_{0}^{t} e^{-\nu k^2(1-\cos(\theta))(t-\tau)}R d\tau\right|=	$$
	
		$$\label{eq:eq2}
		\left| TK\vec{V}_0\right|+
	\left| \int\limits_{ |k-l|>1}\int_{0}^{t} e^{-\nu k^2(1-\cos(\theta))(t-\tau)}\frac{ |R |}{k-l}  k dk	d\tau  \right| +\left| \int\limits_{ |k-l|<1 }\int_{0}^{t} e^{-\nu k^2(1-\cos(\theta))(t-\tau)}\frac{ R }{k-l}   	
	k dkd\tau   \right|
		$$
		$$\label{eq:eq2}
	I_0+I_1+I_2;
	$$
	$$
	I_1<\int_{0}^{t} \frac{C_0\tau}{{(p\nu(1-\cos(\theta))(1-\tau)) }^{2/p}}||\nabla v||_{L_2(R^3)}||v||_{L_2(R^3)}d\tau.
	$$
	$$
I_2=\int_{0}^{t} \int\limits_{ |k-l|<1 }\frac{e^{-\beta^2k^2}( R(k)-R(l) +R(l) )}{k-l}k dk  d\tau=
$$	
	$$
\int_{0}^{t} \int\limits_{ |k-l|<1 }e^{-\beta^2k^2}R^{'}(\gamma)k dk +
R(l)\int_{0}^{t} \int\limits_{ |k-l|<1 }\frac {e^{-\beta^2k^2}  }{k-l}k dk\le
\int_{0}^{t} \left( \frac{|R^{'}|}{\beta} + \frac {|R(l)|}  {\beta}\right) d\tau\le$$
$$
\int_{0}^{t} \frac{1}{\beta} \left(\int\limits_{R^3}|x||v\nabla v|dx + ||\nabla v||_{L_2(R^3)}||v||_{L_2(R^3)}  \right) d\tau\le $$$$\frac{t}{\sqrt{(1-\cos(\theta))}}  \max\limits_{ 0\le \tau<t } \left(\int\limits_{R^3}|x||v\nabla v|dx + ||\nabla v||_{L_2(R^3)}||v||_{L_2(R^3)}  \right)  
 $$	

\end{proof}
%%%%%%%%%%%%%%%%%%%%%%%%%%%%%%%%%%%%%%%%%%%%%%%%%%%%%%%%%%%%%%%%%%%%%

%%%%%%%%%%%%%%%%%%%%%%%%%%%%%%%%%%%%%%%%%%%%%%%%%%%%%%%%%%%%%%%%%%%%%

%%%%%%%%%%%%%%%%%%%%%%%%%%%%%%%%%%%%%%%%%%%%%%%%%%%%%%%%%%%%%%%%%%%%%%%%%%%%%%%%%%%%%%%%%%%%%%%%%
\begin{theorem}
	\textbf{\label{Theorem 8.6}} 
	Let $\mu\vec{v_0} \in W_{2}^{2}(R^{3}),$ $\mu\vec{f}\in L_2(Q_T)$, $\vec{\tilde{f}}\in W_2^{2,1}(Q_T)$, $ \left| TKV_0\right| +\left| TK^2V_0\vec{\tilde{v_0}}\right| <C$,\,\,\,\,$\int_{0}^{\infty}|\mu|  H_0\vec{f}||_{L_{2}(R^{3})}dt<C $. Then,
	the solution of (\ref{1})--(\ref{3}) in Theorem \ref{thm1} satisfies the following inequalities:

	$$
\sup\limits_{0\le\tau<T }	\left\vert \mu\left\vert\nabla 
	\vec{v}\right\vert \right\vert
	_{L_{2}(R^{3})}+
	\nu\int\limits_{0}^{T}\int\limits_{R^{3}}\mu|H_{0} \vec{v}
	|^{2}dxd\tau \leq {\rm const}. 
	$$
		$$
	\sup\limits_{x\in R^{3}}||\vec{v}(x)||<C,
	$$
	
\end{theorem}
\begin{proof}
	
	Consider the Cauchy problem for the Navier--Stokes equations:
	\begin{equation}
	\label{1a}
	%q_{t}-\nu \Delta q+\sum\limits_{k=1}^{3}q_{k}q_{x_{k}}=-\nabla
	\frac{\partial\vec{v}}{\partial t}-\nu \Delta \vec{v}+(\vec{v},\nabla \vec{v})=-\nabla
	p+\vec{f}(x,t),~{\rm div}~\vec{v}=0,  
	\end{equation}
	\begin{equation}\label{2a}
	\vec{v}|_{t=0}=\vec{v}_{0}(x)  
	\end{equation}
	in the domain $Q_{T}=R^{3}\times (0,T)$, where 
	\begin{equation}
	{\rm div}\;\vec{v}_{0}=0.  \label{3a}
	\end{equation}
	We perform the following transformations:
	$$
	\vec{u_{\epsilon}}=\epsilon\vec{v}, \;p_{\epsilon}=p\epsilon,\,\,f_{\epsilon}=f{\epsilon^2},\,\,\nu_{\epsilon}=\epsilon\nu, s=\frac{t}{\epsilon}.
	$$
	Then, 
	\begin{equation}
	\label{1a}
	%q_{t}-\nu \Delta q+\sum\limits_{k=1}^{3}q_{k}q_{x_{k}}=-\nabla
	\frac{\partial\vec{u_{\epsilon}}}{\partial s}-\nu_{\epsilon} \Delta \vec{u_{\epsilon}}+(\vec{u_{\epsilon}},\nabla \vec{u_{\epsilon}})=-\nabla_{\epsilon}
	p_{\epsilon}+\vec{f_{\epsilon}}(x,t),~{\rm div}~\vec{u_{\epsilon}}=0,  
	\end{equation}
	\begin{equation}\label{2a}
	\vec{u_{\epsilon}}|_{t=0}=\vec{u_{\epsilon}}_{0}(x)  
	\end{equation}
	in the domain $Q_{T}=R^{3}\times (0, T_{\epsilon})$, where 
	\begin{equation}
	{\rm div}\;\vec{u_{\epsilon}}|_{t=0}=0.  \label{3a}
	\end{equation}
	Let us return for convenience to the notation $v_{i}=u_{\epsilon_{i}}$,
	using the equation for each   $v_{i}=u_{\epsilon_{i}} $. This gives us
	$$
	-\Delta_{x}\Psi  +v_{i}\Psi  =k^{2}\Psi, ~k\in C.  
	\label{eq:se}
	$$
	Using Lemmas 12-15,  we get estimates for 
	$$ 
	A_{i},\;\vec{V}_{i},\; TA_{i},\;T\vec{V}_{i},\; kA_{i},k\vec{V}_{i}, \;TKA_{i},\;TK\vec{V}_{i},\;TK\tilde{v_{i}},\;TK^2V\tilde{v_{i}}. $$
	The last estimations yield the representation 	
	$$ 
	q= \frac{\Lambda\left(  H_0\int_{S^{2}}\Psi d\theta+k^2\int_{S^{2}}\Psi d\theta\right)  }{\Lambda\int_{S^{2}}\Psi d\theta}|_{r=\frac{\pi}{k_0} ,k=k_0},
	$$
	and   Lemma \ref{lm:l6}	implies	
	$$\label{eq:eq2}
	\left\vert\mu \left\vert\nabla 
	\vec{v}\right\vert \right\vert^2
	_{L_{2}(R^{3})}+\nu_{\epsilon} \int\limits_{0}^{s}||\mu H_{0} \vec{v}||_{L_{2}(R^{3})}^{2}d\tau \le \left\vert \left\vert\nabla 
\mu	\vec{v_{0}}\right\vert \right\vert^2
	_{L_{2}(R^{3})}+   \int_{0}^{s}|| ( \mu\vec{v})||_{L_{2}(R^{3})}|| ||  \mu H_0\vec{f}||_{L_{2}(R^{3})}d\tau    $$
	$$	
	+ \frac{C_0}{\nu_{\epsilon}}\int_{0}^{s}  \max_{x\in R^{3}} |\vec{v}|^2 ||(\mu\nabla \vec{v})||^{2}_{L_{2}(R^{3})}d\tau .
	$$
		$$\label{eq:eq2}
	\left\vert \left\vert\mu \nabla 
	\vec{v}\right\vert \right\vert^2
	_{L_{2}(R^{3})}+\nu_{\epsilon} \int\limits_{0}^{s}||\mu H_{0} \vec{v}||_{L_{2}(R^{3})}^{2}d\tau \le \left\vert \left\vert\mu\nabla 
	\vec{v_{0}}\right\vert \right\vert^2
	_{L_{2}(R^{3})}+   \int_{0}^{s}|| ( \mu\vec{v})||_{L_{2}(R^{3})}|| || \mu H_0\vec{f}||_{L_{2}(R^{3})}d\tau    $$
	$$	
	+ \frac{sC_0}{\nu_{\epsilon}}\int_{0}^{s} \left(\frac{C_1}{\nu_{\epsilon}} ||(\mu\nabla \vec{v})||^{2}_{L_{2}(R^{3})}||( \vec{v})||^{2}_{L_{2}(R^{3})} +|| \mu\vec{v}||^{2}_{L_{2}(R^{3})}\right) ||(\mu\nabla \vec{v})||^{2}_{L_{2}(R^{3})}d\tau .
	$$

	Denote $$
	\alpha(s)=\frac{sC_0}{\nu_{\epsilon}} \left(\frac{C_1}{\nu_{\epsilon}} ||(\mu\nabla \vec{v})||^{2}_{L_{2}(R^{3})}||(\mu \vec{v})||^{2}_{L_{2}(R^{3})} +|| \mu\vec{v}||^{2}_{L_{2}(R^{3})}\right) ,
	$$	
	
	$$\int_{0}^{ \frac{T}{\epsilon}}\alpha(s) ds \le \int_{0}^{T }\frac{TC_0}{\epsilon\nu_{\epsilon}} \left(\frac{C_1}{\nu_{\epsilon}} ||(\mu\nabla \vec{v})||^{2}_{L_{2}(R^{3})}||(\mu \vec{v})||^{2}_{L_{2}(R^{3})} +|| \mu\vec{v}||^{2}_{L_{2}(R^{3})}\right)dt
	$$
	$$
	\le \frac{TC_0C_1}{\epsilon\nu^3_{\epsilon}}\sup\limits_{ t} || ( \mu\vec{v})||^{2}_{L_{2}(R^{3})}\int_{0}^{T}\nu_{\epsilon}||(\mu\nabla \vec{v})||^{2}_{L_{2}(R^{3})}||dt +\frac{C_0}{\nu_{\epsilon}}\sup\limits_{ t} || ( \mu\vec{v})||^{2}_{L_{2}(R^{3})}
	$$
	$$
	\le \frac{TC_0  \epsilon^4}{\epsilon\nu^3_{\epsilon}}	+
	\frac{C_0\epsilon^2} {\nu_{\epsilon}}\le  \frac{2C_0T}{\nu}=C_2 	
	$$

	the Gronwall--Bellman  lemma  yields  
	$$
	\left\vert \left\vert\mu\nabla  
	\vec{v}\right\vert \right\vert^2
	_{L_{2}(R^{3})}+
	\nu_{\epsilon}\int\limits_{0}^{T}\int\limits_{R^{3}}|\mu H_{0} \vec{v}
	|^{2}dxd\tau \leq  	\left\vert \left\vert\nabla 
	\vec{v_0}\right\vert \right\vert^2_ {L_{2}(R^{3})} e^{C_2}$$$$+
	e^{2C_0}\int_{0}^{T}|| ( \mu \vec{v})||_{L_{2}(R^{3})}|| ||\mu  H_0\vec{f}||_{L_{2}(R^{3})}d\tau .
	$$

\end{proof}
%%%%%%%%%%%%%%%%%%%%%%%%%%%%%%%%%%%%%%%%%%%%%%%%%%%%%%%%%%%%%%%%%%%%%
Theorem \ref{Theorem 8.6}  \,\,\,asserts the global solvability and uniqueness of the Cauchy problem for the Navier--Stokes equations.

\section{Discussion}
As noted in the introduction, the key method of investigating the Cauchy problem for the Navier--Stokes equations is its reduction to the Poincar\'e--Riemann--Hilbert problem. By studying the wave functions for the Schr\"dinger equation of the generated velocity components, we obtain unique estimates for the maximum velocity.
Uniform global estimations of the Fourier transform of solutions of the Navier--Stokes equations indicate that the principle modelling of complex flows and related calculations can be based on the Fourier transform method. In terms of the Fourier transform, under both smooth initial conditions and right-hand sides, no exacerbations appear in the speed and pressure modes. A loss of smoothness in terms of the Fourier transform can only be expected in the case of singular initial conditions or of unlimited forces in $ L_ {2} (Q_ {T}) $.
The theory developed by us is supported by numerical calculations performed in Refs. [18--20],
where the dependence of the smoothness of the solution on the oscillations of the system is clearly deduced.

	% %\end{multicols}
	\end{document}